# Optical Characterizations and Electronic Devices of Nearly Pure (10,5) Single-Walled Carbon Nanotubes

Li Zhang[†], Xiaomin Tu[‡], Kevin Welsher[†], Xinran Wang[†], Ming Zheng[‡] and Hongjie Dai[*,†]

[†]*Department of Chemistry and Laboratory for Advanced Materials, Stanford University, Stanford, CA 94305*

[‡]*DuPont Central Research and Development, Wilmington, DE 19880*

Single-walled carbon nanotubes (SWNTs) have attracted phenomenal attention because of their potential applications in nanoelectronics,[1] drug delivery,[2] and biosensors[3]. The properties of SWNTs depend strongly on tube chirality and diameter. Large diameter semiconducting tubes are preferred for nanoelectronics since they form good contact with the electrodes and can carry high current.[4] Recently, chromatography methods have been reported to separate small SWNTs (≤0.9nm) according to their diameter, chirality and length. Unfortunately, the efficiency of this method decreased with increasing tube diameter when using the established ssDNA sequence d(GT)$_n$ (n=10-45).[5-10] The performances of the field effect transistors (FETs) made with these tubes were limited due to their small diameter,[9] demonstrating the need for better chirality separation of large diameter SWNTs for high performance nanoelectronics.

Here we report ion-exchange (IEX) separation of the *(10,5)* SWNT (diameter = 1.03nm) for the application of nanoelectronic devices. A systematic search of DNA sequences for improving IEX separation of DNA-CNT hybrids was performed. This led to the identification of a set of ssDNA sequences, each of which, when used to form hybrids with HiPco tubes, allows particular *(n,m)* semiconducting tubes to elute first among all species during an IEX run, resulting in purification of the desired *(n,m)* species. All 12 major semiconducting tubes present in the starting HiPco material were purified this way. A detailed account of the work will be given elsewhere (Tu et al., manuscript in preparation). The *(10, 5)* tube had the largest diameter among all 12 purified tubes, and was thus chosen for FET device fabrication. It was obtained with the 13-mer ssDNA sequence (TTTA)$_3$T. In addition to the electrostatic properties of the DNA-CNT hybrid, which serves as the foundation for the IEX purification method, we propose that the unique recognition ability of (TTTA)$_3$T for *(10, 5)* tube allows them to form a structurally well-defined complex and elute in early fractions.

The IEX procedure is similar to what was previously reported.[8] However, it was found that the resolution and recovery of each *(n,m)* species can be improved by tuning the composition and pH of the dispersion and elution buffers. For optimum *(10,5)* purification, 1mg ssDNA is added to 1mg HiPCO tubes (Lot R0217, CNI, Houston, TX) in 1mL 0.1M sodium acetate buffer (pH 4.5) and incubated for 2 days before centrifugation. The supernatant solution collected after centrifugation was then fractionated through an IEX column (Biochrom, Terre Haute, IN) with 2xSSC (0.3M NaCl, 0.03M sodium citrate) /0.5mM EDTA /pH 7.0 and a 0-1M sodium benzoate gradient.

The AFM image (Figure 1a) illustrates the purity and length distribution of the separated tubes. They all have the same height, suggesting uniform diameter. The UV-Vis spectrum in Figure 1b shows the $E_{11}$ transition at 1278nm, $E_{22}$ transition at 798nm and

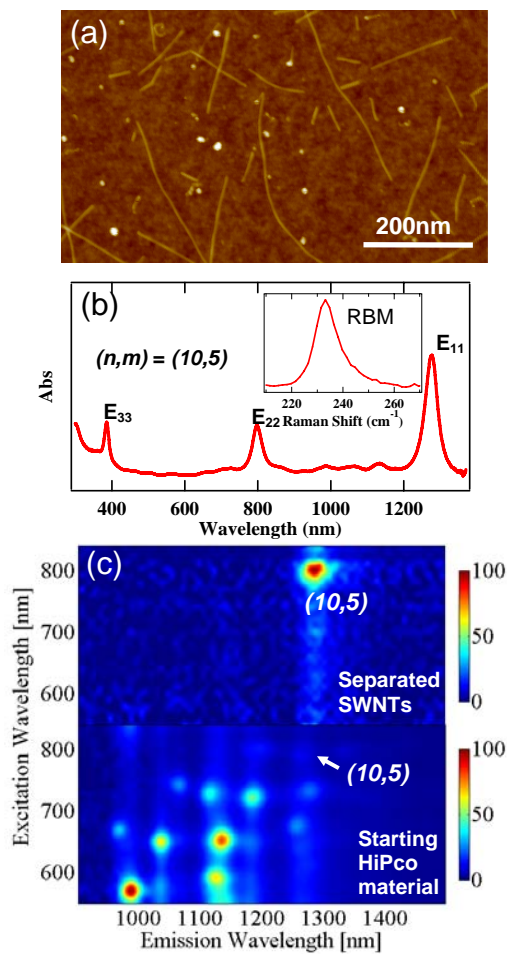

*Figure 1.* (a) AFM image of the *(10,5)* SWNTs, (b) Optical absorption spectrum of the *(10,5)* SWNTs (inset: Raman spectrum), and (c) photoluminescence/excitation spectroscopy (PLE) of the *(10,5)* SWNTs (top) and the starting HiPco material (bottom).

$E_{33}$ transition at 386nm, with $E_{11}$ and $E_{22}$ slightly red shifted from the semi-empirical values (1250nm, 786nm) for *(10,5)* tubes suspended in SDS.[11] The lack of absorption peaks in the 400-600nm region indicates that the amount of metallic SWNTs present in the sample is small and below the UV-Vis detection limit. The resonance Raman spectrum shown in the inset of Figure 1b is the average of spectra collected from hundreds of tubes deposited on silicon substrate. Only one RBM peak at 233cm$^{-1}$ is shown, corresponding to the *(10,5)* tubes. The photoluminescence versus excitation spectra (PLE) shown in Figure 1c provides further information about the purity of the tubes. PLE can clearly distinguish similar diameter semiconducting SWNTs with different chiralities contributing to the same $E_{11}$ absorption peak. The single peak in the spectrum further confirms the single chirality property of the sample. From Figure 1c, one can see that there are few *(10,5)* tubes in the starting HiPco material. Other starting material such as laser ablation tubes are needed to obtain separated large quantities of *(10,5)* SWNTs and other larger SWNTs.

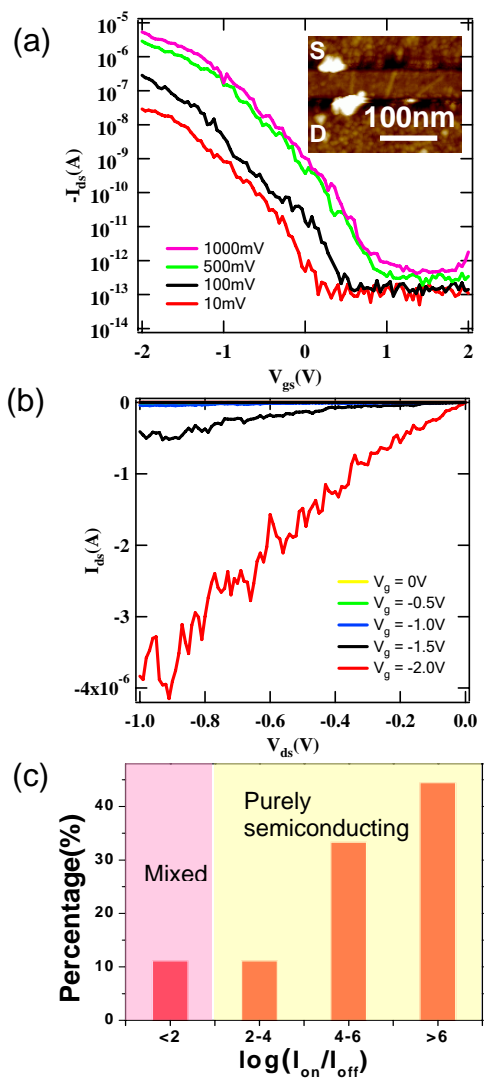

*Figure 2.* Electrical transport properties of FETs made of chemically separated SWNTs. (a) Transfer characteristics ($I_{ds}$-$V_{gs}$ curves) of a typical device made of ~15 *(10,5)* SWNTs at different bias voltages. The inset shows the AFM image of part of the device (b) Current-voltage characteristics ($I_{ds}$-$V_{ds}$) of the device at different $V_{gs}$. (c) Histograms of device percentages with various $I_{on}/I_{off}$ ratios (x-axis), total number of devices is 60.

The electrical properties of the FETs made with the chromatographically separated *(10,5)* SWNTs are shown in Figure 2. The channel length of the devices is around 80-100nm. AFM imaging of the devices shows that each device is connected by ~15 tubes on average (Figure 2a inset). The $I_{on}/I_{off}$ ratio of the typical device is as high as $10^6$ (Figure 2a). Figure 2b shows that the on current at 1.0V bias is ~4μA, which is significantly higher than our previously reported result on the device (~1μA) with longer channel (~200-300nm) and smaller tubes (~0.8-0.9nm).[9] The performance improvement is mainly due to the larger diameter of the SWNTs due to their smaller band gap and smaller contact barriers with the electrodes. The statistic $I_{on}/I_{off}$ of more than 60 devices shows that half of the devices have $I_{on}/I_{off}$ ratio of higher than $10^6$ (Figure 2c). Around 88% of the devices exhibit $I_{on}/I_{off} >10^2$, indicating all SWNTs (average number ~15) in each of these devices are semiconductors (Figure 2c). This suggests that ~ 99% (i.e., $0.99^{15}$ ~ 88%) of the SWNTs in the separated SWNTs are semiconductors, of which almost all are *(10,5)* tubes, according to the PLE spectrum (Figure 1c). In other words, semiconducting *(10,5)* SWNTs are highly enriched, leading to depletable FET devices comprised of SWNTs with the same chirality. The 99% enrichment of *(10,5)* tubes is highly improved over the previous result of 94% enrichment of *(7,6)* and *(8,4)* tubes using the $d(GT)_{20}$ DNA.[9] This is the first time that SWNT FETs with single chirality and diameter ≥1nm have been achieved.

In summary, SWNTs with specific chirality *(10,5)* were recognized by the ssDNA sequence $(TTTA)_3T$ and eluted first during IEX. The separation efficiency was highly improved compared with previously reported results using $d(GT)_{20}$. The separation result was characterized with UV-Vis, Raman and PLE, all of which revealed nearly pure *(10,5)* SWNTs. Nanoelectronic devices were fabricated with $I_{on}/I_{off}$ as high as $10^6$, owning to the single chirality enriched *(10,5)* tubes. While the performance of the FETs made with the slightly bigger *(10,5)* tubes are better than the previously reported result using *(7,6)* and *(8,4)* tubes, even larger single chirality semiconducting SWNTs are preferred to further improve the device performance. We have identified the sequence motif that has the potential to purify single-chirality SWNTs. Applying these sequences to laser ablation SWNTs may purify other large diameter tubes.

**Acknowledgment**. This work was supported by MARCO-MSD, Intel and NSF Grant CMS-060950.

**Supporting Information Available:** The details of the photoluminescence excitation measurement, Raman, AFM, and electrical device fabrication and measurement are described. This material is available free of charge via the Internet at http://pubs.acs.org.

TOC

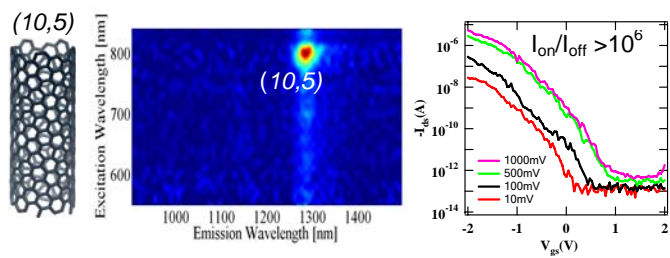

ABSTRACT

It remains an elusive goal to achieve high performance single-walled carbon nanotube (SWNTs) field effect transistors (FETs) comprised of only single chirality SWNTs. Many separation mechanisms have been devised and various degrees of separation demonstrated, yet it is still difficult to reach the goal of total fractionation of a given nanotube mixture into its single chirality components. Chromatography has been reported to separate small SWNTs (diameter $\leq$ 0.9nm) according to their diameter, chirality and length. The separation efficiency decreased with increasing tube diameter by using ssDNA sequence d(GT)$_{n\ (n=10-45)}$. Here we report our result on the separation of single chirality *(10,5)* SWNTs (diameter = 1.03nm) from HiPco tubes with ion exchange chromatography. The separation efficiency was improved by using a new DNA sequence (TTTA)$_3$T which can recognize SWNTs with the specific chirality *(10,5)*. The chirality of the separated tubes was examined by optical absorption, Raman, photoluminescence excitation/emission and electrical transport measurement. All spectroscopic methods gave single peak of *(10,5)* tubes. The purity was 99% according to the electrical measurement. The FETs comprised of separated SWNTs in parallel gave $I_{on}/I_{off}$ ratio up to $10^6$ owning to the single chirality enriched *(10,5)* tubes. This is the first time that SWNT FETs with single chirality SWNTs were achieved. The chromatography method has the potential to separate even larger diameter semiconducting SWNTs from other starting material for further improving the performance of the SWNT FETs.